\documentclass[12pt]{article}
\pdfoutput=1

\usepackage{draft} 
\usepackage{hyperref}
\usepackage{graphicx,color}
\usepackage{cite}
\usepackage{mciteplus}
\usepackage{skak}

\begin{document}

\unitlength = .8mm

\begin{titlepage}
\rightline{MIT-CTP-4644}

\begin{center}

\hfill \\
\hfill \\
\vskip 1cm

\title{Constraining Higher Derivative Supergravity
\\ with Scattering Amplitudes}

\author{Yifan Wang$^\textsymknight$ and Xi Yin$^\textsymbishop$}

\address{$^\textsymknight$Center for Theoretical Physics, Massachusetts Institute of Technology, \\
Cambridge, MA 02139 USA
\\
$^\textsymbishop$Jefferson Physical Laboratory, Harvard University, \\
Cambridge, MA 02138 USA}

\email{yifanw@mit.edu,
xiyin@fas.harvard.edu}

\end{center}

\abstract{ We study supersymmetry constraints on higher derivative deformations of type IIB supergravity by consideration of superamplitudes. Combining constraints of on-shell supervertices and basic results from string perturbation theory, we give a simple argument for the non-renormalization theorem of Green and Sethi, and some of its generalizations.
}

\vfill

\end{titlepage}

\eject


%

Supersymmetry constraints on higher derivative couplings in maximal supergravity theories have been investigated extensively in the past \cite{Green:1997tv,Berkovits:1997pj,Green:1998by,Pioline:1998mn,Berkovits:1998ex,Green:2005ba,Basu:2008cf} and have led to remarkable exact results on the quantum effective action of string theory. The method of 
\cite{Paban:1998ea,Paban:1998qy,Sethi:1999qv,Green:1998by} in obtaining constraints on higher derivative terms in gauge and gravity theories with maximal supersymmetry was by explicitly analyzing supersymmetry variations of fields and the Lagrangian and their deformations, starting from the purely fermionic terms. In this note we present a simple argument for such non-renormalization theorems from scattering amplitudes, in the context of ten dimensional type IIB supergravity and its deformations, largely inspired by work of \cite{Elvang:2010xn,Elvang:2010jv,Elvang:2013cua} on the classification of supergravity counter terms \cite{Kallosh:1980fi,Howe:1980th,Howe:1981xy} using amplitudes.

To begin with, we recall the spinor helicity formulation of superamplitudes in type IIB supergravity \cite{CaronHuot:2010rj,Boels:2012ie}. A 10 dimensional null momentum $p^m$ and the corresponding (constrained) spinor helicity variables $\lambda_A^\A$ are related by 
\ie
p^m \delta_{AB} = \Gamma^m_{\A\B} \lambda_A^\A \lambda_B^\B,
\fe
where $\A$ is a chiral spinor index of $SO(1,9)$ and $A$ is a spinor index of the $SO(8)$ little group. The $2^8=256$ states in the supergraviton multiplet are built from monomials in a set of Grassmann variables $\eta_A$. The supermomentum is then defined as
\ie
q^\A = \lambda^\A_A \eta^A.
\fe
A typical $n$-point superamplitude takes the form\footnote
{
The cubic vertex is special \cite{Boels:2012ie}, and may be constructed as follows. Define $\widehat N^\A{}_\B = p_1^m p_2^n (\Gamma_{mn})^\A{}_\B$ which specifies the null plane spanned by the three external momenta $p_1,p_2,p_3$. We have $\widehat N\lambda_{iA}=0$ for $i=1,2,3$. We can label the $SO(1,9)$ spinor components by $(s_0s_1s_2s_3s_4)$, $s_a=\pm$, such that $\widehat N=(p^+)^2 \Gamma^{0^-1^-}$, where $\Gamma^{0^-1^-}$ is the lowering operator on both $s_0$ and $s_1$. Now decompose the spinor helicity variables according to their $s_0s_1$ spin, $\lambda_{iA} = (\lambda_{iA}^{++},\lambda_{iA}^{+-},\lambda_{iA}^{-+},\lambda_{iA}^{--})$, where each $\lambda_{iA}^{\pm\pm}$ is a spinor of the $SO(6)$ tiny group that acts transversely to the null plane, and the condition $\widehat N\lambda_{iA}=0$ amounts to $\lambda_{iA}^{++}=0$. Note define the tiny group spinor valued supermomentum $W^{\pm\pm} = \sum_{i}^3 \lambda_{iA}^{\pm\pm} \eta_i^A$. The cubic supervertex is then given by the boost invariant combination ${1\over (p^+)^4} \delta^4(W^{+-})\delta^4(W^{-+})\delta^4(W^{--})$.
}
\ie
{\cal A} = \delta^{10} (P) \delta^{16}(Q) {\cal F}(\lambda_i,\eta_i),
\fe
where $P=\sum_{i=1}^n p_i$, and the 32 supercharges that act on the $n$-particle asymptotic states can be expressed as
\ie
Q^\A = \sum_{i=1}^n q_i^\A,~~~~ \widetilde Q^\A = \sum_{i=1}^n \lambda_i^{\A A} {\partial\over \partial \eta_i^A}.
\fe
They obey $\{Q^\A, \widetilde Q^\B\} = {1\over 2}  \Gamma_m^{\A\B} P^m$.
The nontrivial supersymmetry Ward identities on ${\cal A}$ are
\ie
\delta^{10} \left( P \right) \delta^{16}(Q) \,\widetilde Q^\A\Big[ {\cal F}(\lambda_i,\eta_i) \Big] = 0.
\fe
We can write the CPT conjugate of the amplitude ${\cal A}$ as
\ie
\overline{\cal A} = \delta^{10} (P) \widetilde Q^{16} {\cal F}(\lambda_i,\partial/\partial\eta_i) \prod_{i=1}^n \eta_i^8.
\fe
Evidently, if ${\cal A}$ obeys supersymmetry Ward identities, so does $\overline{\cal A}$.

Now let us focus on supervertices, namely superamplitudes with no poles in momenta. There are three basic types of supervertices we can write down. First, we can take ${\cal F}(\lambda_i,\eta_i)$ to be independent of $\eta_i$, namely
\ie
{\cal F}(\lambda_i,\eta_i) = f(s_{ij}),
\fe
where $s_{ij}=-(p_i+p_j)^2 = -2p_i\cdot p_j$. The CPT conjugate of this construction gives another supervertex. 
We refer to these as F-term vertices.\footnote{These vertices are also known as ``Maximal $R$-symmetry violating" (MRV) in \cite{Boels:2012zr,Boels:2013jua}.}
A third type of supervertex (D-term) is given by
\ie
\delta^{10}(P) \delta^{16}(Q)\,\widetilde Q^{16} h(\lambda_i,\eta_i).
\fe
Here $h$ is an arbitrary function of the spinor helicity variables.
All supervertices we know of are of these three types. We conjecture that these are in fact the only supervertices that obey supersymmetry Ward identities, and will proceed with this assumption.

Let us inspect a particularly simple set of $n=(4+k)$-point F-term vertices, with ${\cal F}(\lambda_i,\eta_i)=1$,
\ie
\delta^{10}(P)\delta^{16}(Q).
\fe
In component fields, we will expand the axion-dilaton field as $\tau=\tau_0+\varphi$, where $\tau_0$ is the background value. Such a vertex then corresponds to an independent set of couplings in the Lagrangian of the form\cite{Berkovits:1997pj,Pioline:1998mn}
\ie
\varphi^k R^4 + \cdots.
\fe
Similarly, the conjugate vertex 
\ie
\delta^{10}(P) \widetilde Q^{16} \prod_{i=1}^{4+k}\eta_i^8
\fe
corresponds to the coupling $\overline\varphi^k R^4+\cdots$. Note that in the $k=0$ case, $\delta^{16}(Q) = \widetilde Q^{16} \prod_{i=1}^4 \eta_i^8$ is self-conjugate, and corresponds to the $R^4$ vertex.\footnote{In contrast, the supergravity 4-point tree amplitude is given by ${\delta^{10}(P)\delta^{16}(Q) \over stu}$ \cite{CaronHuot:2010rj,Boels:2012ie}.} In particular, we see that there are {\it no independent} supervertex of the form $\varphi^k \overline\varphi^\ell R^4 + \cdots$ with $k,\ell\geq 1$.

Note that in a superamplitude, two $SO(8)$ little group invariant monomials in $\eta_i^A$, namely 1 and $\eta_i^8$, correspond to the $i$-th external particle being $\varphi$ and $\overline\varphi$ respectively. The nonlinearly realized $SL(2,\mathbb{R})$ of type IIB supergravity is broken by the expectation value of $\tau$ to a $U(1)$,\footnote{While this $U(1)$ acts on the target space of the axion-dilaton field locally as an isometry, in type IIB string theory it is incompatible with the $SL(2,\mathbb{Z})$ identification.} which acts on the amplitude by $\sum_i \big({1\over 4} \eta_i {\partial\over \partial\eta_i} - 1 \big)$ and assign opposite charges to $\varphi$ and $\overline\varphi$. This $SL(2,\mathbb{R})$ is generally broken explicitly by the higher derivative supervertices of consideration here.

Now, we would like to constrain the coupling
\ie
f(\tau,\bar\tau) R^4 + \cdots
\fe
by type IIB supersymmetry. In a vacuum in which $\tau$ acquires constant expectation value $\tau_0$, expanding $\tau= \tau_0 + \varphi$, we obtain a series of operators,
\ie
f(\tau_0,\bar\tau_0) R^4 + \partial_\tau f(\tau_0,\bar\tau_0) \varphi R^4 + \partial_{\overline \tau} f(\tau_0,\bar\tau_0) \overline\varphi R^4 + \partial_\tau \partial_{\overline \tau} f(\tau_0,\bar\tau_0) \varphi \overline\varphi R^4 +\cdots
\fe
Since there are independent $\varphi R^4$ and $\overline\varphi R^4$ supervertices, $\partial_\tau f$ and $\partial_{\bar\tau} f$ can take arbitrary value at $\tau=\tau_0$. This reflects a freedom in adjusting $f(\tau,\bar\tau)$ by a holomorphic and an anti-holomorphic function of $\tau$. $\partial_\tau \partial_{\overline \tau} f$ at $\tau=\tau_0$, on the other hand, is not independent, because there is no independent $\varphi\overline\varphi R^4$ vertex. This 6-point coupling therefore must be constrained in terms of the $R^4$ coefficient, namely $f(\tau_0,\bar\tau_0)$, by supersymmetry.

\begin{figure}[htb]
\centering
\begin{minipage}{0.5\textwidth}
\centering
\includegraphics[scale=1.3]{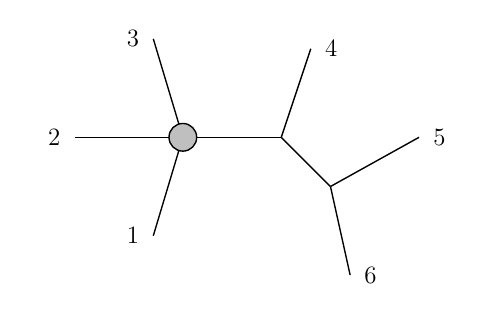}
\end{minipage}\hfill
\begin{minipage}{0.5\textwidth}
\centering
\includegraphics[scale=1.3]{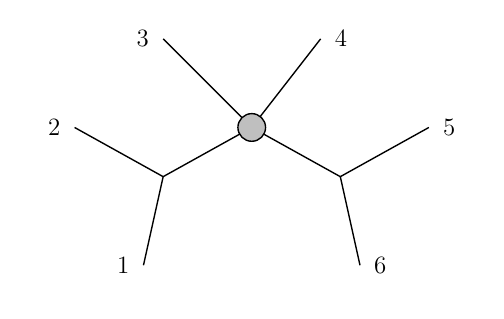}
\end{minipage}
\caption{Factorization of the $6$-point amplitude through one $D^n R^4$ vertex and a pair of supergravity cubic vertices.}
\label{sewing1}
\end{figure}

In principle, one can ask for the most general 6-point superamplitude that obeys supersymmetry Ward identities and factorization through lower point amplitudes by unitarity. By dimension analysis, the 6-point $\varphi$-$\overline\varphi$-$R^4$ superamplitude could only factorize through a single $R^4$ supervertex and supergravity vertices (Figure \ref{sewing1}). The $\varphi\overline\varphi R^4$ coupling itself can then be recovered by taking the soft limit on a pair of $\varphi$ and $\overline\varphi$ scalar lines \cite{Cheung:2014dqa}.
 We do not know a systematic way of building higher point superamplitudes with the $R^4$ on-shell supervertex.\footnote{For instance, if one applies BCFW \cite{Britto:2005fq} shift to a pair of external lines and try to rewrite the higher point tree amplitude as a contour integral in the shift parameter $z$, one encounters nontrivial residue at $z=\infty$, which cannot be determined in a straightforward way. The all-line shift of \cite{Cohen:2010mi} improves the behavior at $z=\infty$ but still does not appear to apply when general higher derivative vertices are present.} However, from unitarity we know that such a relation must exist, and is linear in this case, namely
\ie
({\rm Im}\tau_0)^2 \partial_\tau \partial_{\overline \tau} f(\tau_0,\bar\tau_0) \propto f(\tau_0,\bar\tau_0),
\fe
where the $({\rm Im}\tau_0)^2$ factor comes from the normalization of the dilaton-axion kinetic term. To determine the relative coefficient, it suffices to find {\it any} set of such couplings that solve the supersymmetry and unitarity constraints. String perturbation theory already gives such a solution. Since the tree level effective action of type IIB string theory contains $R^4$ coupling at $\alpha'^3$ order, it suffices to examine this coupling in Einstein frame, which takes the form
\ie
\tau_2^{3/ 2} R^4.
\fe
Since $\partial_\tau \partial_{\overline\tau} \tau_2^{3/ 2} = {3\over 16} \tau_2^{-1/ 2}$, we immediately obtain the relation
\ie
4 ({\rm Im}\tau_0)^2 \partial_\tau \partial_{\overline \tau} f(\tau_0,\bar\tau_0) = {3\over 4}  f(\tau_0,\bar\tau_0),
\fe
which must then hold for the general $f(\tau,\bar\tau)$ at all values of $\tau_0$. This is the non-renormalization theorem of Green and Sethi \cite{Green:1998by}. In below, we will write $f_n(\tau,\bar\tau)$ for the coefficient of $D^n R^4$, and so $f(\tau,\bar\tau)$ will be denoted $f_0(\tau,\bar\tau)$.

\begin{figure}[htb]
\centering
\includegraphics[scale=1.3]{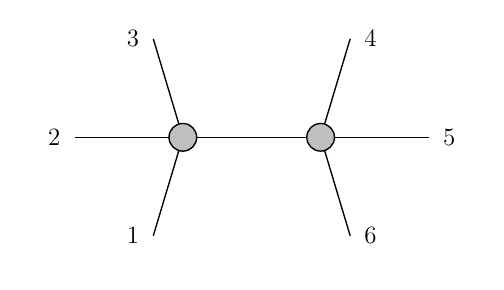}
\caption{Factorization of the $6$-point amplitude though a pair of $R^4$ vertices.}
\label{sewing2}
\end{figure}

Note that there is no independent $D^2 R^4$ supervertex, as the corresponding superamplitude must be proportional to $\delta^{16}(Q)(s+t+u)=0$. We next apply the argument to $f_4(\tau,\bar\tau) D^4 R^4$ coupling. Once again, the holomorphic and anti-holomorphic parts of $f_4(\tau,\bar\tau)$ are unconstrained by supersymmetry, as there are independent $\varphi^k R^4$ and $\overline\varphi^k R^4$ supervertices. $\partial_\tau \partial_{\overline\tau} f_4$, on the other hand, must obey a linear relation with $\tau_2^{-2} f_4(\tau,\bar\tau)$, due to the factorization of 6-point superamplitude. Note that the 6-point amplitude at this order in the momentum expansion does not factorize through two $R^4$ vertices (Figure \ref{sewing2}), as the latter can only contribute to the 6-point amplitude at $D^6 R^4$ order\footnote{
This can be seen from the corresponding BCFW \cite{Boels:2012ie} residues: for the factorization in Figure \ref{sewing2}, it takes the form
\ie
{\delta^{16}(Q) \over s_{123}} \int d^8\eta_P \delta^{16}(q_P+q_4+q_5+q_6).
\fe 
}. 

Now taking the IIB string tree level effective action, and expanding to $\A'^5$ order, we find in Einstein frame the coupling
\ie
\tau_2^{5/ 2} (s^2+t^2+u^2) R^4.
\fe
By comparison, we then immediately obtain the relation
\ie
4 \tau_2^2 \partial_\tau \partial_{\overline \tau} f_4(\tau,\bar\tau) = {15\over 4} f_4(\tau,\bar\tau).
\fe

At $f_6(\tau, \bar\tau) D^6 R^4$ order, we encounter a novelty: as already mentioned, the 6-point amplitude at this order in the momentum expansion admits a factorization into a pair of $R^4$ supervertices.
Thus, we expect the coefficient $f_6(\tau,\bar\tau)$ to obey a relation of the form
\ie\label{abr}
\tau_2^2 \partial_\tau \partial_{\overline\tau} f_6 = a  f_6(\tau,\bar\tau) + b f_0(\tau,\bar\tau)^2,
\fe
where $a,b$ are two constants. More precisely, we define $f_6(\tau,\bar\tau)$ to be the coefficient of $(s^3+t^3+u^3) R^4 = 3 stu R^4$. Inspecting the well-known string tree level massless 4-point amplitude, 
\ie
&  \delta^{16}(Q) {\Gamma(-{\A' s\over 4}) \Gamma(-{\A' t\over 4}) \Gamma(-{\A' u\over 4})\over \Gamma(1+{\A' s\over 4}) \Gamma(1+{\A' t\over 4}) \Gamma(1+{\A' u\over 4})}
\\
&=  \delta^{16}(Q) \left[ - {64\over \A'^3 stu} - 2 \zeta(3) - {\zeta(5)\over 16} \A'^2 (s^2+t^2+u^2) - {\zeta(3)^2\over 96}\A'^3 (s^3+t^3+u^3) 
+ \cdots  \right] ,
\fe
we can identify the following couplings in Einstein frame,\footnote{Effective action couplings from higher-point superstring tree amplitudes have also been extracted, for example, in \cite{Barreiro:2012aw,Barreiro:2013dpa} for type I open strings and in \cite{Basu:2013goa} for type II closed strings.}
\ie
-2 \zeta(3) \tau_2^{3/2} \A'^3 R^4 - {\zeta(5)\over 16} \A'^5 \tau_2^{5/2} (s^2+t^2+u^2) R^4
- {\zeta(3)^2\over 96} \A'^6 \tau_2^3 (s^3+t^3+u^3) R^4  
+ \cdots
\fe
Comparing to (\ref{abr}), with $f_0\propto \tau_2^{3/2}$ and $f_6\propto \tau_2^3$, we immediately obtain a linear relation between $a$ and $b$.
Another relation between $a$ and $b$ may be extracted from the string 1-loop effective action. The perturbative contribution to $f_0$ and $f_6$ can be expanded in the form \cite{Green:2005ba}
\ie
f_n(\tau,\bar\tau) = f_n^{tree} + f_n^{1-loop} + f_n^{2-loop}+f_n^{3-loop}+\cdots.
\fe
In particular, at 1-loop order, we expect
\ie
\tau_2^2 \partial_\tau \partial_{\overline\tau} f_6^{1-loop} = a  f_6^{1-loop}(\tau,\bar\tau) + 2 b f_0^{tree}(\tau,\bar\tau) f_0^{1-loop}(\tau,\bar\tau).
\fe
The 4-point massless genus one string amplitude has analytic as well as non-analytic terms in the momentum expansion. The $R^4$ term, with coefficient $f_0^{1-loop}\propto \tau_2^{-1/2}$, and the $D^6 R^4$ term, with coefficient $f_6^{1-loop}\propto \tau_2$, are analytic, and were computed in \cite{Green:1999pv}. They give an independent linear relation which then fixes $a$ and $b$, as in (5.39) of \cite{Green:2005ba}. In the end, one finds
\ie
4 \tau_2^2 \partial_\tau \partial_{\overline\tau} f_6 = 12 f_6(\tau,\bar\tau) - 6 f_0(\tau,\bar\tau)^2.
\fe
As was pointed out in \cite{Green:2005ba}, the string 3-loop contribution $f_6^{3-loop}$ \cite{Green:2005ba,Gomez:2013sla,D'Hoker:2014gfa,Basu:2014hsa,Pioline:2015yea}, proportional to $\tau_2^{-3}$, is what solves the homogeneous version of the constraining equation (namely, it is annihilated by $4\tau_2^2\partial_\tau \partial_{\bar\tau} - 12$).

Now let us consider $D^8 R^4$ terms. There is again one independent 4-point supervertex one can write down,\footnote{Note that $(s^2+t^2+u^2)^2$ is proportional to $(s^4+t^4+u^4)$, with $s+t+u=0$.}
\ie\label{fds}
\delta^{16}(Q) (s^4+t^4+u^4).
\fe
This is in fact proportional to the D-term vertex
\ie
\delta^{16}(Q)\, \widetilde Q^{16} \left[ \sum_{i<j}^4 \eta_i^8 \eta_j^8 \right].
\fe
To understand the constraints on $f_8(\tau,\bar\tau)$, let us inspect $(n=4+k)$-point supervertices of the form
\ie
\delta^{16}(Q)\, \widetilde Q^{16} F(\eta_i^8),
\fe
where $F(\eta_i^8)$ is a polynomial in the little group invariants $\eta_i^8$, of total degree $8m$ in the $\eta$'s, for some integer $m\geq 2$. This then corresponds to a coupling of the form $\varphi^{k-m+2} \overline\varphi^{m-2} D^8 R^4$. Since these D-term vertices by construction obey supersymmetry Ward identities, there are no constraint on the coefficients of $\varphi^{k-m+2} \overline\varphi^{m-2} D^8 R^4$, thus no constraint on $f_8(\tau,\bar\tau)$ from supersymmetry alone.

At order $D^{10} R^4$, there is again just one independent 4-point supervertex $\delta^{16}(Q) (s^5+t^5+u^5)$. This is proportional to the D-term vertex $\delta^{16}(Q)\, \widetilde Q^{16} \left[ \sum_{i<j}^n s_{ij} \eta_i^8 \eta_j^8 \right]$.\footnote{One may try to write down another D-term vertex using little group invariants, of the form $
\delta^{16}(Q)\, \widetilde Q^{16} \big[ (\eta_1^7\lambda_1)^\A (\eta_2^7\lambda_2)^\B q_3^\C q_4^\D (\Gamma^{mnp})_{\A\B} (\Gamma_{mnp})_{\C\D} + {\rm permutations} \big]$.
Nonetheless, this must be proportional to $\delta^{16}(Q)\, \widetilde Q^{16} \left[ \sum_{i<j} s_{ij} \eta_i^8 \eta_j^8 \right]$ and cannot be an independent vertex.}
As in the $D^8R^4$ case, there are no supersymmetry constraints on the coefficient $f_{10}(\tau,\bar{\tau})$. In other words, the differential constraint proposed in \cite{Basu:2006cs} should be a consequence of additional properties in IIB string theory.

In conclusion, the formulation of higher derivative couplings in maximally supersymmetric gravity theories in terms of on-shell supervertices gives a simple classification of independent couplings allowed by supersymmetry. When combined with solutions to supersymmetry Ward identities provided by string perturbation theory, the consideration of supervertices then leads to a derivation of type IIB supersymmetry constraints on the F-term $f(\tau,\bar\tau) D^n R^4$ coupling. The result is nonetheless a consequence of maximal supersymmetry on higher derivative supergravity theories, and no longer depend on string theory. Clearly, this strategy generalizes to maximal supergravity theories in other dimensions as well.\footnote{Except for type IIA and eleven dimensional supergravity, where a supermomentum formulation of the amplitude as presented here is unavailable.}

Finally, let us comment on the role of $SL(2,\mathbb{R})$ symmetry of type IIB supergravity which, as already mentioned, is explicitly broken by these higher derivative terms. A coupling of the form $f_n(\tau,\bar\tau) D^n R^4$ violates $SL(2,\mathbb{R})$ unless $f_n$ is a constant, but the latter is incompatible with the supersymmetry constraints (a nontrivial second order differential equation in $\tau,\bar\tau$) for F-term vertices. From this perspective, a role of the nonlinearly realized $SL(2,\mathbb{R})$ symmetry of type IIB supergravity is to rule out F-terms as potential counter terms. Indeed, the UV divergence in type IIB supergravity first arises at two-loop order, corresponding to an $SL(2,\mathbb{R})$-invariant D-term counter term of the form $D^{10} R^4$. One may expect that the $E_{7(7)}$ symmetry of four dimensional maximal supergravity plays a similar role in that it rules out F-terms as counter terms, but there appear to be plenty of D-term supervertices that are compatible with $E_{7(7)}$ that could serve as counter terms \cite{Brodel:2009hu,Beisert:2010jx,Bossard:2010bd,Kallosh:2014hga,Bern:1998ug,Bern:2006kd,Bern:2007hh,Bern:2009kd,Bern:2011qn}. 

\bigskip
\bigskip

We would like thank Nima Arkani-Hamed, Clay Cordova, Thomas Dumitrescu, Henriette Elvang, Daniel Freedman, Simeon Hellerman, Yu-tin Huang, Hermann Nicolai, Alexander Zhiboedov, and especially Zohar Komargodski for extensive discussions. We are grateful to the hospitality of Weizmann institute, Jerusalem Winter School, and Kavli IPMU during the course of this work.  Y.W. is supported in part by the U.S. Department of Energy under grant Contract Number  DE-SC00012567. X.Y. is supported by a Sloan Fellowship and a Simons Investigator Award from the Simons Foundation. 

\bibliography{IIBrefs} 
\bibliographystyle{JHEP}
\end{document}